\documentclass[12pt]{article}

\usepackage{amssymb,amsmath,graphics,hyperref}

\newcommand{\RR}{{\mathbb{R}}}

\newcommand{\pa}{\partial}
\newcommand{\ii}{{\rm i}}
\newcommand{\dd}{{\rm d}}

\newcommand{\sfrac}[2]{{\textstyle\frac{#1}{#2}}}
\newcommand{\Tr}{\mathrm{Tr}\,}
\newcommand{\Vg}{\mathrm{dV}_g}
\newcommand{\Vh}{\mathrm{dV}_h}

\title{Topological energy bounds for the Skyrme and Faddeev models with massive pions}
\author{Derek Harland\footnote{email address: d.g.harland@leeds.ac.uk}
  \bigskip
  \\School of Mathematics,
  \\University of Leeds,
  \\LS2 9JT}
\date{8th November 2013}

\begin{document}

\maketitle

\begin{abstract}
 A topological lower bound on the Skyrme energy which depends explicity on the pion mass is derived.
 This bound coincides with the previously best known bound when the pion mass vanishes, and improves on it whenever the pion mass is non-zero.  The new bound can in particular circumstances be saturated.  New energy bounds are also derived for the Skyrme model on a compact manifold, for the Faddeev-Skyrme model with a potential term, and for the Aratyn-Ferreira-Zimerman and Nicole models.
\end{abstract}

\section{Introduction}
\label{sec1}

The Skyrme model \cite{skyrme} is a model of atomic nuclei in which the baryon number is identified with a topological invariant, and nuclei appear as topological solitons called skyrmions.  This model appears as an effective description of QCD in the limit of a large number of colours \cite{witten}.  Despite possessing only a small number of parameters, the model successfully captures many properties of nuclei, including their spectra of excited states \cite{bmsw}.

A key feature of the Skyrme model is the topological energy bound \cite{faddeev76}.  This states that the energy $E$ of any configuration with baryon number $B$ is greater than a positive constant $C$ times $|B|$, thereby encapsulating the idea that masses of nuclei are roughly proportional to their baryon numbers.

Topological energy bounds provide insight into binding energies.  If the scaling law $E=C|B|$ is satisfied exactly by minimal-energy solitons then binding energies are zero, since solitons with baryon number $B$ can break up into solitons of lower charge at no energetic cost.  Similarly, if soliton energies $E_B$ are only slightly larger than $C|B|$ then the binding energies are small, since the difference $E_{B_1}+E_{B_2}-E_{B_1+B_2}$ can be no greater than $\sum_{n=1}^2(E_{B_n}-C|B_n|)$.  Thus one way to obtain realistically small binding energies is to design a model with a topological energy bound which is almost, but not quite, saturated.  This idea is at the heart of various extensions \cite{sutcliffe,as-gw} of the Skyrme model proposed in the last few years.

Another recent development in the Skyrme model has been the inclusion of a pion mass term.  Including this term has led to more realistic spatial energy distributions \cite{bs}, and has also revealed a link with the alpha-particle model of nuclei \cite{bms}.  Studies \cite{bmsw,bks} of excited states based on semiclassical quantisation indicate that the optimum value for the bare pion mass is somewhat larger than the physically observed pion mass; it is presumed that this bare mass would be renormalised to a lower value in a full quantisation of the model.

In this note a topological energy bound will be presented for the Skyrme model with pion mass term.  This bound depends on the value of the pion mass and is stronger than the standard bound \cite{faddeev76} whenever the pion mass is non-zero.  Moreover, comparison with numerical data indicates that skyrmions with massive pions come closer to saturating their lower bound than those with massless pions.  In fact, in a particular limit the new bound is exactly saturated.  All of this suggests that the pion mass term favours low binding energies.

The same types of arguments used to derive this new bound can also be applied to a variant of the Skyrme model introduced by Faddeev \cite{faddeev}.  Again, a new mass-dependent bound can be derived which improves on the standard bound \cite{vk} whenever the mass term is non-zero.  As a by-product, topological energy bounds are obtained for the variants of the Faddeev-Skyrme model proposed by Nicole \cite{nicole} and Aratyn-Ferreira-Zimerman \cite{afz}.  Applied to the Skyrme model on a compact domain, our arguments yield a topological energy bound which scales as $B^{4/3}$ (a bound for the Faddeev-Skyrme model on a compact domain has previously been obtained in \cite{ss}).

The new bound for the Skyrme model will be derived and analysed in sections \ref{sec2} and \ref{sec3}, and the bound for Faddeev's model and its variants will be derived in section \ref{sec4}.  Some conclusions will be drawn in section \ref{sec5}.

\section{The extreme Skyrme model}
\label{sec2}

\subsection{A lower bound}

The matter content of the Skyrme model is a map $\phi:M\to\Sigma$, where $(M,g)$ and $(\Sigma,h)$ are both three-dimensional Riemannian manifolds.  In typical applications to nuclear physics one takes $M=\RR^3$ and $\Sigma=S^3\cong {\rm SU}(2)$.  The strain tensor for $\phi$ is by definition $D_i^j = g^{jk} \pa_i\phi^\alpha\pa_k\phi^\beta h_{\alpha\beta}$.  Skyrme's energy functional is \cite{manton}
\begin{equation}
\label{Skyrme energy 1}
 E = \int_M \big[ \alpha_2\Tr D + \alpha_4\sfrac12((\Tr D)^2-\Tr(D^2)) + \alpha_0 V(\phi)\big]\Vg ,
\end{equation}
where $\alpha_0,\alpha_2,\alpha_4$ are non-negative real parameters, $V$ is a non-negative real function on $\Sigma$, and $\Vg=\sqrt{g}\dd^3x$ is the Riemannian volume form on $M$.  The eigenvalues of $D$ are non-negative and will be denoted $\lambda_1^2,\lambda_2^2,\lambda_3^2$, and the first two terms in the energy density can be reexpressed as $\Tr D=\sum_i\lambda_i^2$ and $\sfrac12((\Tr D)^2-\Tr(D^2))=\sum_{i<j}\lambda_i^2\lambda_j^2$.

If both $M$ and $\Sigma$ are compact without boundary, the map $\phi$ has a topological invariant $B\in\mathbb{Z}$, known as the degree or the topological charge.  The degree may be computed using the formula,
\begin{equation}
\label{degree formula}
 \int_M\phi^\ast\Omega = B\int_\Sigma\Omega,
\end{equation}
in which $\Omega$ is any volume form on $\Sigma$.  If $M$ is not compact, the degree is still well-defined provided that the condition that $\phi$ is constant on the boundary of $M$ is imposed.  For example, if $M= \RR^3$ and $\Sigma={\rm SU}(2)$ it is required that $\phi(\mathbf{x})$ tends to the identity matrix as $r\to\infty$, and the integer $B$ is identified with the baryon number in this case.

Faddeev derived a lower bound on the first two terms in the energy functional \cite{faddeev76}:
\begin{equation}
\label{Faddeev LB}
 E \geq 6\sqrt{\alpha_2\alpha_4}\,Vol(\Sigma)\,|B|.
\end{equation}
Below, a bound will be derived on the second and third terms in the energy functional; this will be combined with Faddeev's bound in the section that follows.  Accordingly, we set $\alpha_2=0$ and without loss of generality assume that $\alpha_4=\alpha_0=1$; then the energy is
\begin{equation}
 E = \int_M \left[ \lambda_1^2\lambda_2^2 + \lambda_1^2\lambda_3^2 + \lambda_2^2\lambda_3^2 + V(\phi) \right]\Vg.
\end{equation}

The main tool in the derivation of the bound is the inequality of the arithmetic and geometric means: if $w_a$ are $n$ positive real numbers that sum to 1 and $x_a$ are $n$ non-negative real numbers, then
\begin{equation}
\label{AMGM}
 \sum_{a=1}^n w_a x_a \geq \prod_{a=1}^n x_a^{w_a}
\end{equation}
with equality if and only if $x_1=x_2=\ldots=x_n$.  We also make use of H\"older's inequality,
\begin{equation}
 \label{Holder}
 \left(\int_{M}|f_1|^p\mathrm{dV}_g\right)^{\frac1p} \left(\int_{M}|f_2|^q\mathrm{dV}_g\right)^{\frac1q} \geq \int_M |f_1f_2| \mathrm{dV}_g,
\end{equation}
valid whenever $1/p+1/q=1$ and $f_1,f_2$ are functions such that the left hand side is finite.  Equality holds in this expression if and only if one of the functions $f_i$ is equal to a constant times the other.

The first application of the inequality \eqref{AMGM} yields
\begin{equation}
\label{ES1}
 E\geq 4\left( \frac13\int_M \left[\lambda_1^2\lambda_2^2 + \lambda_1^2\lambda_3^2 + \lambda_2^2\lambda_3^2\right]\Vg\right)^{3/4} 
 \left( \int_M V(\phi) \Vg \right)^{1/4}.
\end{equation}
Here the two weights $w_a$ have been chosen to be $\sfrac14$ and $\sfrac34$ so that the expression on the right is scale invariant when $M=\RR^3$.  If the weights had not been so chosen the right hand side would be unstable to scalings, and in particular not bounded from below by any positive number.

The next step uses the inequality \eqref{AMGM} again to deduce that
\begin{equation}
\label{ES2}
\frac13 (\lambda_1^2\lambda_2^2 + \lambda_1^2\lambda_3^2 + \lambda_2^2\lambda_3^2) \geq |\lambda_1\lambda_2\lambda_3|^{4/3}.
\end{equation}
From H\"older's inequality it follows that
\begin{equation}
\label{ES3}
 4\left( \int_M |\lambda_1\lambda_2\lambda_3|^{4/3} \Vg \right)^{3/4} \left( \int_M V(\phi) \Vg\right)^{1/4} 
 \geq 4 \int_M V^{1/4}|\lambda_1\lambda_2\lambda_3|\Vg.
\end{equation}
The quantity on the right of this expression is greater than or equal to the integral over $\RR^3$ of $\phi^\ast(V^{1/4}\Vh)$, as follows from the identity $\det D^2\det g=\det(\pa\phi^\alpha/\pa x^i)^2\det h$.  Thus by equation \eqref{degree formula} the following bound holds:
\begin{equation}
\label{extreme Skyrme bound}
 E \geq 4 |B| \int_\Sigma V^{1/4}\Vh.
\end{equation}

\subsection{Saturating the bound}

It is instructive to consider whether the bound \eqref{extreme Skyrme bound} can be saturated.  The first inequality \eqref{ES1} in the derivation is saturated if and only if
\begin{equation}
\label{ES4}
 \int_M \left[\lambda_1^2\lambda_2^2 + \lambda_1^2\lambda_3^2 + \lambda_2^2\lambda_3^2\right] \Vg = 3\int_M V(\phi) \Vg.
\end{equation}
It is noteworthy that when $M=\RR^3$ this equation is precisely the condition that $\phi$ is stable to Derrick scalings $\phi(x)\mapsto \phi(\lambda x)$, so this condition is satisfied by any (finite-energy) solution of the field equations.

The second inequality \eqref{ES4} holds if and only if $\lambda_1=\lambda_2=\lambda_3$.  This condition is equivalent to the statement that $\phi^\ast h = \lambda^2 g$ for some real function $\lambda=\lambda_i$, and in particular is true when $\phi$ is a conformal map.  The third inequality \eqref{ES3} holds if and only if $V(\phi(x)) = C\lambda^4(x)$ for some positive real constant $C$.  From equation \eqref{ES4} it is clear that $C=1$.  Therefore the bound \eqref{extreme Skyrme bound} is saturated if and only if $\phi$ is a map such that
\begin{equation}
\label{ES saturation}
 \phi^\ast h = \sqrt{V\circ\phi}\, g.
\end{equation}

There are certainly maps which satisfy equation \eqref{ES saturation}.  For example, let $\phi:\RR^3\to S^3$ be the inverse stereographic projection,
\begin{equation}
 \phi(x^1,x^2,x^3) = \left( \frac{1-|{\bf x}|^2}{1+|{\bf x}|^2}, \frac{2x^1}{1+|{\bf x}|^2}, \frac{2x^2}{1+|{\bf x}|^2}, \frac{2x^3}{1+|{\bf x}|^2} \right).
\end{equation}
This map is a conformal, and the pull-back of the metric on the sphere is $\phi^\ast h = \left(2/(1+|{\bf x}|^2)\right)^2\dd x^i\dd x^i$.
Thus equation \eqref{ES saturation} is satisfied by this map for the particular choice of potential,
\begin{equation}
\label{ES potential}
 V(\phi^0,\phi^1,\phi^2,\phi^3) = (1+\phi^0)^4.
\end{equation}
The energy of this map is equal to its lower bound \eqref{extreme Skyrme bound}, and takes the numerical value $E=8\pi^2$.  In contrast, Faddeev's bound \eqref{Faddeev LB} for the standard Skyrme model is saturated only by isometries, and thus can never be attained by maps from $\RR^3$ to $S^3$.

For the particular choice of potential \eqref{ES potential} the bound \eqref{extreme Skyrme bound} can be saturated only when $|B|=0$ or 1.  Thus the energy of any skyrmion of topological charge $B>1$ is greater than $B$ times the energy of the 1-skyrmion.  In the standard Skyrme model with potential $V=1+\phi^0$ the energies of $B$-skyrmions are significantly less than $B$ times the energy of the 1-skyrmion.  The behaviour or real nuclei lies somewhere between these two extremes, with the mass of a nucleus with baryon number $B$ being only slightly less than $B$ times the proton mass.  Thus one might hope that physically realistic binding energies could be achieved in the Skyrme model by a judicious choice of potential function.

\subsection{The Skyrme model on a compact manifold}

When $M$ has finite volume and $|B|$ is large the bound \eqref{extreme Skyrme bound} is not the strongest possible: a stronger bound can be obtained from the Skyrme term alone.  Accordingly, let us assume that $\alpha_2=\alpha_0=0$ and $\alpha_4=1$, and write
\begin{equation}
 E = \int_M \left( \lambda_1^2\lambda_2^2 + \lambda_1^2\lambda_3^2 + \lambda_2^2\lambda_3^2 \right)\Vg.
\end{equation}
H\"older's inequality imples that
\begin{equation}
 \left( \int_M |\lambda_1\lambda_2\lambda_3|^{\sfrac43} \Vg\right)^{\sfrac34} 
 \left( \int_M \Vg\right)^{\sfrac14} 
\geq \int_M |\lambda_1\lambda_2\lambda_3|  \Vg.
\end{equation}
The quantity on the right of this inequality is greater than or equal to the integral over $M$ of the pull-back of the volume form on $\Sigma$.  Thus by equation \eqref{degree formula} the quantity on the right is greater than or equal to $B$ times the volume of $\Sigma$.  In view of the inequality \eqref{ES2} the bound
\begin{equation}
\label{FV lower bound}
 E \geq 3|B|^{\sfrac43}\frac{ {\rm Vol}(\Sigma)^{\sfrac43} }{ {\rm Vol}(M)^{\sfrac13} }
\end{equation}
is obtained.  This inequality is saturated if and only if $\phi$ is an isometry up to scale, that is, $\phi^\ast h = C g$ for some constant $C$.  Clearly the lower bound \eqref{FV lower bound} still applies (with an additional factor of $\alpha_4$) to the more general energy functional \eqref{Skyrme energy 1}.  Since this bound is proportional to $|B|^{4/3}$ rather than $|B|$, it exceeds Faddeev's bound and the bound \eqref{extreme Skyrme bound} for large enough $|B|$.

Skyrme models on compact manifolds $M$ (such as the three-torus) are used as models of nuclear matter at high density \cite{cjjvj,ks}.  The bound \eqref{FV lower bound} should have some relevance there; indeed, the special case $B=1$ of this bound was previously derived by Manton \cite{manton} in this context.

\section{The standard Skyrme model}
\label{sec3}

In the present section the bound \eqref{extreme Skyrme bound} will be combined with Faddeev's bound to yield a lower bound which is stronger than either.  Attention will now be restricted to the case $M=\RR^3$ and $\Sigma={\rm SU}(2)\cong S^3$; accordingly, the Skyrme field will be an SU(2)-valued function $U({\bf x})$.  In standard units, Skyrme's energy functional is
\begin{equation}
 E = \frac{F_\pi^2}{8} E_2 + \frac{1}{2e^2} E_4 + \frac{m_\pi^2F_\pi^2}{8}E_0,
\end{equation}
where
\begin{align}
 E_2 &= \int_{\RR^3} -\frac{1}{2}\Tr(R_iR_i) \,\dd^3x, \\
 E_4 &= \int_{\RR^3} - \frac{1}{16}\Tr([R_i,R_j][R_i,R_j]) \,\dd^3 x, \\
 E_0 &= \int_{\RR^3}\Tr(1-U) \,\dd^3x,
\end{align}
and $R_i=\pa_i UU^{-1}$.  Faddeev's lower bound \eqref{Faddeev LB} is
\begin{equation}
\label{Faddeev}
 \alpha_2E_2+\alpha_4E_4 \geq 12\pi^2 \alpha_2^{1/2}\alpha_4^{1/2}|B|,
\end{equation}
and the lower bound \eqref{extreme Skyrme bound} is
\begin{equation}
\label{me}
 \alpha_0E_0+\alpha_4E_4\geq 16\pi I \alpha_0^{1/4}\alpha_4^{3/4} |B|,
\end{equation}
where
\begin{equation}
 I = \int_0^\pi (2(1-\cos\theta))^{\sfrac14}\sin^2\theta\dd\theta\approx 1.807.
\end{equation}

The idea pursued in this section is to split the energy functional into two pieces and apply the two bounds \eqref{Faddeev} and \eqref{me} simultaneously.  Thus let $t\in[0,1]$ be a parameter and write
\begin{align}
 E &= \left( \frac{F_\pi^2}{8} E_2 + \frac{1-t}{2e^2} E_4 \right) + \left( \frac{m_\pi^2F_\pi^2}{8}E_0 + \frac{t}{2e^2}E_4\right) \\
& \geq 12\pi^2 \frac{F_\pi}{4e}(1-t)^{1/2}|B| + 16\pi I\left(\frac{m_\pi F_\pi}{8e^3}\right)^\sfrac12 t^{3/4} |B| \\
\label{SS1}
& = \frac{12\pi^2 F_\pi |B|}{4e} \left( (1-t)^{1/2} + \frac{2}{3}\sqrt{\frac{\mu}{2}} t^{3/4} \right),
\end{align}
where in the last line the dimensionless parameter
\begin{equation}
 \mu = \frac{16 I^2m_\pi}{\pi^2 F_\pi e}
\end{equation}
has been introduced for notational convenience.  The lower bound \eqref{SS1} is a function of $t$ that attains its maximum when $t = \mu/(1 + \sqrt{1+\mu^2})$.  Thus the strongest lower bound attainable by the above method is
\begin{equation}
\label{SS lower bound}
 E \geq 12\pi^2 |B|\, \frac{F_\pi}{4e} \left(1+\frac{1}{3}\frac{\mu^2}{1+\sqrt{1+\mu^2}} \right)\left(\frac{2}{1+\sqrt{1+\mu^2}}\right)^{\sfrac12}.
\end{equation}
When $\mu=0$ this is just Faddeev's bound \eqref{Faddeev}, while in the limit $\mu\to\infty$ this is the lower bound \eqref{me}.  For all intermediate values of $\mu$ the bound is stronger than either.  Currently in applications to nuclear physics the most popular choice of parameters has $m:=2m_\pi/eF_\pi= 1$ \cite{bmsw,bks}.  With this value, the combined bound \eqref{SS lower bound} is 16\% above Faddeev's bound and 52\% above the lower bound \eqref{me}.

It is informative to compare the bound \eqref{SS lower bound} with skyrmion energies quoted in the literature.  In \cite{bms} skyrmions are constructed numerically in the model with $m=1$ with topological charges in the range $4\leq B\leq 32$.  The energies are between 28\% and 30\% above Faddeev's bound, and hence between 10\% and 12\% above the bound \eqref{SS lower bound}.  In \cite{bs} it was noted that Skyrme energies scale like $\sqrt{m_\pi}$ as $m_\pi\to\infty$ with $e$ and $F_\pi$ fixed; the bound \eqref{SS lower bound} exhibits similar scaling behaviour.

\begin{figure}
\centering
\includegraphics{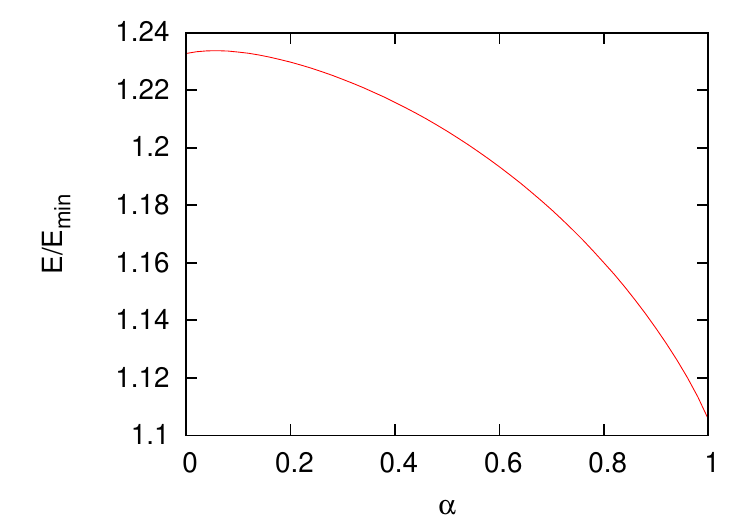}
\caption{Energy of a 1-skyrmion as a function of $\alpha$.  $E_{min}$ is the lower bound \eqref{SS lower bound}.}
\label{fig}
\end{figure}

Numerical simulations indicate that the minimal energy charge 1 skyrmion is spherically symmetric.  We have compared the energy of the spherically-symmetric 1-skyrmion with the lower bound for a large range of values of $m$ using the following standard procedure.  First, a spherically-symmetric hedgehog ansatz is made for the Skyrme field:
\begin{equation}
 U(\mathbf{x}) = \exp(\ii f(r) \sigma_j x^j/r).
\end{equation}
The boundary conditions $f(0)=\pi$, $f(r)\to 0$ as $r\to\infty$ are imposed on the real function $f$ so that the topological charge of $U$ is 1.  The units of length and energy are chosen so that Skyrme's energy is $(1-\alpha) E_2 + E_4 + \alpha E_0$.  Doing so gives the skyrmion a roughly constant size and energy as $\alpha$ is varied, so that the same numerical grid can be used for all values of $\alpha$.  With these units, $m=\sqrt{\alpha}/(1-\alpha)$ and $\mu=8I^2m/\pi^2$.  Substitution of the ansatz into the energy functional yields
\begin{multline}
 E = 4\pi \int_0^\infty \bigg[ (1-\alpha)\left((f')^2+2\frac{\sin^2f}{r^2}\right)  \\
+ \frac{\sin^2f}{r^2}\left(2(f')^2+\frac{\sin^2f}{r^2}\right) + 2\alpha(1-\cos f) \bigg] r^2\dd r.
\end{multline}
A discretised version of this energy with first order derivatives was minimised using an annealing method.  The number of gridpoints and the stepsize were chosen so that doubling either did not significantly alter the energies.  The resulting energies are plotted as a function of $\alpha$ in figure \ref{fig}.  The excess of the soliton energy above its lower bound decreases from 23\% at $m=0$ to 11\% at $m=\infty$; apart from a slight rise near $m=0$, the decrease is monotonic.  This supports the hypothesis that increasing the size of the potential term reduces binding energies.

\section{The Faddeev-Skyrme model}
\label{sec4}

In this section the ideas developed above will be applied to another model that supports soliton solutions, namely the Faddeev-Skyrme model \cite{faddeev}.  The field content of Faddeev's model is a map $\phi:\RR^3\to S^2$.

The map $\phi$ will be written $(\phi^1({\bf x}),\phi^2({\bf x}),\phi^3({\bf x}))$ such that $\vec{\phi}\cdot\vec{\phi}=1$, and the strain tensor is $D_i^j=\pa_i\phi^a\pa^j\phi^a$.  The energy functional is
\begin{equation}
 E = \int_{\RR^3}\left[ \alpha_2\Tr D + \frac{\alpha_4}{2}((\Tr D)^2-\Tr(D^2)) + \alpha_0V(\phi)\right]\dd^3 x ,
\label{SF energy}
\end{equation}
with $\alpha_0$, $\alpha_2$ and $\alpha_4$ non-negative real parameters and $V$ a non-negative real function on $S^2$.  Maps $\phi$ which tends to a constant as $r\to\infty$ can be extended to maps from $S^3$ to $S^2$ and therefore possess a topological invariant $Q\in\pi_3(S^2)\cong\mathbb{Z}$, the Hopf degree.  Vakulenko and Kapitanski obtained a lower bound \cite{vk} on $E$ in terms of the Hopf invariant:
\begin{equation}
\label{VK bound}
 E\geq 3^{3/8}16\pi^2 \sqrt{\alpha_2\alpha_4} |Q|^{3/4}
\end{equation}
(see also \cite{ly}).  It is known that the power $\sfrac34$ of $|Q|$ is optimal \cite{ly}, but it has been conjectured that the coefficient $3^{3/8}16\pi^2$ can be significantly improved \cite{ward}.

\subsection{The bound}

Clearly the Vakulenko-Kapitanski bound makes no reference to the third term $V(\phi)$ in the energy density.  Here a bound will be obtained on the second and third terms, which will subsequently be combined with the Vakulenko-Kapitanski bound.  Thus to begin suppose that $\alpha_2=0$ and $\alpha_4=\alpha_0=1$.  Since the target of $\phi$ is two-dimensional, the strain tensor $D$ has only two non-zero eigenvalues, denoted $\lambda_1^2$ and $\lambda_2^2$, and $\sfrac12((\Tr D)^2-\Tr(D^2))=\lambda_1^2\lambda_2^2$.  The first steps in the derivation of a lower bound mirrors those in the Skyrme model:
\begin{align}
 E &\geq 4\left(\int_{\RR^3}\frac{1}{3}\lambda_1^2\lambda_2^2\dd^3x\right)^{\frac34} \left(\int_{\RR^3}V\dd^3x\right)^{\frac14} \\
 &\geq \frac{4}{3^{3/4}}\int_{\RR^3}V^{1/4}|\lambda_1\lambda_2|^{3/2}\dd^3x \\
\label{AFZ energy}
 &= \frac{4}{3^{3/4}}\int_{\RR^3}(\mathbf{B}.\mathbf{B})^{\sfrac34}\dd^3x.
\end{align}
In the last equality of this sequence, the integrand has been reexpressed in terms of
\begin{equation}
\label{def B}
 B^i=\sfrac12\epsilon^{ijk}V^{1/6}\vec\phi.\pa_j\vec\phi\times\pa_k\vec\phi.
\end{equation}
This $\mathbf{B}$ is the unique vector field such that $i_{\bf B}\dd^3x = \phi^\ast \Omega$, where $\Omega$ is the standard area form on $S^2$ multiplied by $V^{1/6}$ and $i$ denotes the inner derivative.  This vector field is identically divergenceless.

The next part of the derivation relies on the following formula for the Hopf invariant:
\begin{equation}
\label{Hopf degree}
 Q = \frac{1}{\left(\int_{S^2}\Omega\right)^2}\int_{\RR^3\times\RR^3} \frac{\mathbf{B}(x)\times\mathbf{B}(y).(\mathbf{x}-\mathbf{y})}{4\pi|\mathbf{x}-\mathbf{y}|^3}\dd^3x\dd^3 y.
\end{equation}
This formula can be deduced by at least two different methods.  The first method begins with Whitehead's formula for the Hopf invariant as a Chern-Simons (or helicity) integral:
\begin{equation}
 Q = \frac{1}{\left(\int_{S^2}\Omega\right)^2}\int_{\RR^3} \mathbf{A}(x).\mathbf{B}(x)\dd^3x,
\end{equation}
in which $\mathbf{A}$ is a vector potential satisfying $\nabla\times\mathbf{A}=\mathbf{B}$.  A well-known Green's function formula for the gauge potential in Coulomb gauge ($\nabla.\mathbf{A}=0$) is
\begin{equation}
 \mathbf{A}(x) = \frac{1}{4\pi}\int_{\RR^3} \mathbf{B}(y)\times \frac{\mathbf{x}-\mathbf{y}}{|\mathbf{x}-\mathbf{y}|^3} \dd^3y.
\end{equation}
Substitution of this expression into Whitehead's formula yields equation \eqref{Hopf degree}.  This proof relies on the assumption that $\mathbf{B}$ decays fast enough as $r\to \infty$ for the Green's function formula to be valid (and for the gauge potential $\mathbf{A}$ to extend to $S^3$).  A second derivation of equation \eqref{Hopf degree} is based on an interpretation as an average linking number, and appears in the appendix.

Freedman and He have shown \cite{fh}, using the Hardy-Littlewood-Sobolev inequality, that the integral appearing in \eqref{AFZ energy} is bounded from below by the integral \eqref{Hopf degree} representing the Hopf degree:
\begin{multline}
\label{Freedman He}
 \int_{\RR^3}|\mathbf{B}(x)|^{3/2}\dd^3x 
 \geq C \left(\int_{\RR^3\times\RR^3} \frac{\mathbf{B}(x)\times\mathbf{B}(y).(\mathbf{x}-\mathbf{y})}{4\pi|\mathbf{x}-\mathbf{y}|^3}\dd^3x\dd^3 y\right)^{3/4} \\ \mbox{ where } C = \left(\frac{16}{\pi}\right)^{\frac14}.
\end{multline}
Combined with the inequalities preceding \eqref{AFZ energy}, this yields the bound
\begin{equation}
\label{EF lower bound}
 E \geq \frac{8}{(27\pi)^{1/4}} \left(\int_{S^2}\Omega\right)^{\frac32} |Q|^{3/4}.
\end{equation}

\subsection{Comparison with numerical data}

The question now arises as to how close the bound \eqref{EF lower bound} comes to being saturated.  A numerical study of minimisers of the energy functional \eqref{SF energy} was carried out in \cite{foster}, with the particular choice $V=2(1-\phi_3)$ of potential function.  The lower bound \eqref{EF lower bound} for the model with $(\alpha_0,\alpha_2,\alpha_4)=(1,0,1)$ is $E/|Q|^{3/4}\geq2^83^{3/4}\pi^{5/4}7^{-3/2}\approx132$.  The smallest value of $E/|Q|^{3/4}$ obtained in this case was $0.82\times 32\pi^2\sqrt{2}\approx 366$, which is roughly 2.78 times the lower bound.  This is of similar magnitude to the excess of minimisers of the energy with $\alpha_0=0$ above the Vakulenko-Kapitanski bound.

If all three coefficients $\alpha_0$, $\alpha_2$ and $\alpha_4$ are non-zero the lower bound \eqref{EF lower bound} may be used in combination with the Vakulenko-Kapitanski bound \eqref{VK bound}.  The particular combinations of terms contributing to the combined bound can be optimised to obtain the strongest possible bound following the method presented in section \ref{sec3}.  Omitting the details, the final result is
\begin{equation}
\label{SF lower bound}
 E \geq 3^{3/8}16\pi^2 \sqrt{\alpha_2\alpha_4}\,  |Q|^{3/4} 
 \left(1+\frac{1}{3}\frac{\mu^2}{1+\sqrt{1+\mu^2}} \right)\left(\frac{2}{1+\sqrt{1+\mu^2}}\right)^{\sfrac12},
\end{equation}
with
\begin{equation}
 \mu = \frac{2^7 3^{11/4}}{7^{3}\pi^{3/2}} \sqrt{\frac{\alpha_4\alpha_0}{\alpha_2^2}}.
\end{equation}
The minimal energies in the $Q=1$ sector obtained in \cite{foster} are listed in table \ref{table}.
\begin{table}
\centering
 \begin{tabular}{|c|c|}
  \hline
  $m=\sqrt{\alpha_4\alpha_0/\alpha_2^2}$ & $E/E_{min}$ \\
  \hline
  0 & 2.32 \\
  1 & 2.51 \\
  2 & 2.67 \\
  4 & 2.71 \\
  5 & 2.72 \\
  $\infty$ & 2.91 \\
  \hline
 \end{tabular}
\label{table}
\caption{Energies of 1-solitons in the Faddeev-Skyrme model; $E_{min}$ is the lower bound of eq.\ \eqref{SF lower bound}}
\end{table}

\subsection{Alternative energy functionals}

Aratyn, Ferreira and Zimerman introduced \cite{afz} the following variant of the Skyrme-Faddeev energy:
\begin{equation}
 E_{AFZ} = \int_{\RR^3}\left( \sfrac{1}{2} \pa_i\vec\phi\times\pa_j\vec\phi\,.\,\pa_i\vec\phi\times\pa_j\vec\phi\right)^{\frac34} \dd^3 x.
\end{equation}
The integrand in this expression is equal to $|\mathbf{B}|^{3/2}$, where now $B^i=\sfrac12\epsilon^{ijk}\vec\phi.\pa_j\vec\phi\times\pa_k\vec\phi$.  Thus Freedman and He's inequality \eqref{Freedman He} leads directly to a lower bound $E_{AFZ}\geq 16\pi^{5/4}|Q|^{3/4}$.  Minima of $E_{AFZ}$ have been studied both numerically \cite{gillard} and analytically \cite{afz}; the smallest known value of $E/|Q|^{3/4}$ is given by an analytic solution and is equal to $16\pi^2$.  This exceeds the lower bound by a factor of $\pi^{3/4}\approx2.36$.

Nicole studied \cite{nicole} the energy functional,
\begin{equation}
 E_N = \int_{\RR^3} \left( \pa_i\vec\phi. \pa_i\vec\phi \right)^{\frac32}\dd^3 x.
\end{equation}
A straightforward application of the inequality \eqref{AMGM} shows that $E_N\geq 2^{3/2}E_{AFZ}$, and hence that $E_N\geq 32\sqrt{2}\pi^{5/4}|Q|^{3/4}$.  In a comprehensive numerical study \cite{gs} the lowest value of $E_N/|Q|^{3/4}$ was attained by an analytical configuration with $Q=1$; the energy of this configuration is $32\sqrt{2}\pi^2$, which once again is $\pi^{3/4}$ times the lower bound.

\section{Conclusions}
\label{sec5}

In this note a new pion mass-dependent lower bound \eqref{SS lower bound} on the Skyrme energy functional has been derived.  As the pion mass increases this bound becomes more effective, in the sense that the ratios $E/E_{min}$ between soliton energies and their lower bound decrease towards 1.  With a particular choice of potential \eqref{ES potential} and in the absence of an $E_2$ term in the energy, the bound can be saturated exactly.

These results suggest that a Skyrme model with realistically low binding energies might be obtained by judiciously choosing a potential function based on \eqref{ES potential}.  A full investigation of this idea would involve fully three-dimensional simulations of the field equations, which are beyond the scope of the current investigation.

A new bound \eqref{SF lower bound} for the Faddeev-Skyrme model with potential term has also been derived, based on an inequality \eqref{Freedman He} which is in effect a lower bound on the AFZ energy.  Unlike in the Skyrme model, solitons in this model do not come close to saturating the lower bound.  The root of this difficulty seems to be the inequality \eqref{Freedman He}, which is apparently far from optimal.  The lowest-energy solitons in the AFZ model exceed this lower bound by a factor of $\pi^{3/4}$, suggesting that the best value for $C$ in \eqref{Freedman He} is in fact $\sqrt{4\pi}$.  If this were true it would be possible to prove a stronger lower bound for the Faddeev-Skyrme model, even in the case of vanishing potential: indeed, elementary applications of the inequalities \eqref{AMGM}, \eqref{Holder} and \eqref{Freedman He} result in the bound $E_2+E_4\geq (8\pi)^{3/2}C|Q|^{3/4}$.  When $C=\sqrt{4\pi}$ this exceeds the bound \eqref{VK bound}, and in fact coincides with the bound conjectured by Ward \cite{ward}.

The methods used to derive both of these bounds could more widely.  In a forthcoming publication \cite{aw} an energy bound will be derived for a more general Skyrme model that includes higher-derivative terms.

\bigskip\noindent\textbf{Acknowledgements}
I wish to thank Martin Speight for discussions, and D.\ Foster, M.\ Gillard, P.\ Sutcliffe and C.\ Adam for suggesting improvements to a draft of this paper.  I am grateful to the authors of \cite{aw} for sharing their results.

\section*{Appendix: The Hopf degree}

In this appendix an alternative derivation of equation \eqref{Hopf degree} will be supplied.  Recall that if the preimages of two points $u,v\in S^2$ under $\phi$ are differentiable curves in $\RR^3$ parametrised as $\vec\gamma_u(s)$ and $\vec\gamma_v(t)$, then the Hopf invariant is equal to their linking number.  This may be calculated using Gauss' formula:
\begin{equation}
 Lk(\phi^{-1}(u),\phi^{-1}(v)) 
 = \frac{1}{4\pi} \int \frac{\dot\gamma_u(s)\times\dot\gamma_v(t).(\mathbf{\gamma}_u(s)-\mathbf{\gamma}_v(t))}{|\mathbf{\gamma}_u(s)-\mathbf{\gamma}_v(t)|^3}\dd s\dd t.
\end{equation}

For fixed differentiable $\phi$, denote by $D$ the set in $\RR^3$ on which $\dd\phi\neq0$, and let $U$ denote the set of points $u\in S^2$ such that $\phi^{-1}(u)\subset D$.  Then $B=0$ outside of $D$, and the sets $S^2\setminus U$ and $D\setminus\phi^{-1}(U)$ have measure 0.  The preimage of any point $u\in U$ is a differentiable curve.  We suppose that these curves can be parametrised as $\gamma_u(s)$, such that
\begin{equation}
\label{def gamma}
 \dot\gamma_u^i(s)=B^i(\gamma_u(s))
\end{equation}
and such that $\gamma$ is a differentiable bijection from a subset $V\subset U\times \RR$ to $\phi^{-1}(U)$.

It follows immediately from the definitions \eqref{def gamma} of $\gamma$ and \eqref{def B} of $\mathbf{B}$ that the pull-back of the volume form $\dd^3 x$ under $\gamma$ is equal to $\dd s \wedge\Omega$.  Therefore, up to sets of measure zero,
\begin{align}
 &\int_{\RR^3\times\RR^3} \frac{\mathbf{B}(x)\times\mathbf{B}(y).(\mathbf{x}-\mathbf{y})}{4\pi|\mathbf{x}-\mathbf{y}|^3}\dd^3x\dd^3 y \\
&= \int_{V\times V} \frac{\dot\gamma_u(s)\times\dot\gamma_v(t).(\mathbf{\gamma}_u(s)-\mathbf{\gamma}_v(t))}{|\mathbf{\gamma}_u(s)-\mathbf{\gamma}_v(t)|^3}\dd s\,\Omega(u)\,\dd t\,\Omega(v) \\
&= \int_{U\times U} Lk(\phi^{-1}(u)\phi^{-1}(v) \,\Omega(u)\wedge\Omega(v) \\
&= Q \left(\int_{S^2}\Omega\right)^2.
\end{align}
This proves equation \eqref{Hopf degree} under the assumptions on $\phi$ outlined above.  Most of these assumptions can be relaxed.  For topological reasons it may not be possible to define the ``inverse'' $\gamma$ of $\phi$ globally on $U$.  However, $\gamma$ can be defined on local patches in $U$, and the argument goes through with integrals over $U$ replaced by a sum of integrals over these patches.  It may also happen that for some $u\in U$ the inverse image $\phi^{-1}(u)$ consists of a collection of closed curves rather than a single curve.  In this situation one could divide $U$ into regions $U^{(n)}$ in which $\phi^{-1}(u)$ has $n$ components, and then apply the change of variables separately on each of these regions.

\bibliographystyle{latexeu}
\bibliography{skyrmelb}

\end{document}